\documentstyle[12pt]{article}
\begin{document}
%  Greek letters
\def\a{\alpha}
\def\b{\beta}
\def\ch{\chi}
\def\d{\delta}
\def\e{\epsilon}
\def\f{\phi}
\def\g{\gamma}
\def\h{\eta}
\def\i{\iota}
\def\j{\psi}
\def\k{\kappa}
\def\l{\lambda}
\def\m{\mu}
\def\n{\nu}
\def\o{\omega}
\def\p{\pi}
\def\q{\theta}
\def\r{\rho}
\def\s{\sigma}
\def\t{\tau}
\def\u{\upsilon}
\def\x{\xi}
\def\z{\zeta}
\def\D{\Delta}
\def\F{\Phi}
\def\G{\Gamma}
\def\J{\Psi}
\def\L{\Lambda}
\def\O{\Omega}
\def\P{\Pi}
\def\S{\Sigma}
\def\U{\Upsilon}
\def\X{\Xi}
\def\T{\Theta}

\def\Ab{\bar{A}}
\def\gi{g^{-1}}
\def\li{{ 1 \over \l } }
\def\lb{\l^{*}}
\def\zb{\bar{z}}
\def\ub{u^{*}}
\def\Tb{\bar{T}}
 \def\pp {\partial }
\def\pb {\bar{\partial }}
\def\be{\begin{equation}}
\def\ee{\end{equation}}
\def\ben{\begin{eqnarray}}
\def\een{\end{eqnarray}}

\addtolength{\topmargin}{-0.8in}
\addtolength{\textheight}{1in}
\hsize=16.5truecm
\hoffset=-.5in
\baselineskip=7mm

\thispagestyle{empty}
\begin{flushright} \ November \ 1996\\
SNUTP 96-112 \\
\end{flushright}

\begin{center}
 {\large\bf
Self-Dual Chern-Simons Solitons and\\
 Generalized Heisenberg Ferromagnet Models }
\vglue .2in
Phillial Oh\footnote{ E-mail address; ploh@newton.skku.ac.kr }
\vglue .2in
{\it
Department of Physics,
Sung Kyun Kwan University \\
Suwon, 440-746,  Korea}
\vglue .2in
Q-Han Park\footnote{ E-mail address; qpark@nms.kyunghee.ac.kr }
\vglue .2in
{\it Department of Physics, Kyunghee University\\
Seoul, 130-701, Korea}
\vglue .2in
{\bf ABSTRACT}  
\end{center}
We consider the (2+1)-dimensional gauged Heisenberg 
ferromagnet model coupled with the Chern-Simons gauge fields. 
Self-dual Chern-Simons solitons, the static zero energy solution
saturating Bogomol'nyi bounds, are shown to exist when the generalized
spin variable is valued in the Hermitian symmetric spaces $G/H$. 
By gauging the maximal torus subgroup of $H$, we obtain self-dual 
solitons which satisfy vortex-type nonlinear equations thereby 
extending the two dimensional instantons in a nontrivial way. 
An explicit example for the $CP(N)$ case is given.

\vglue .2in

\newpage
Recently, there appeared an action principle of the generalized Heisenberg 
ferromagnet model in terms of a nonrelativistic nonlinear sigma model
defined on a Lie group $G$ \cite{park}. This action possesses a local $H$ subgroup 
symmetry so that the physical spin variables take value on the coadjoint 
orbit of the Hermitian symmetric space $G/H$. The symplectic structure on 
each orbit also allows a direct first order action in terms of generalized 
spin variables. The use of Hermitian symmetric space made possible a systematic 
generalization of the Heisenberg ferromagnet model according to the Cartan's 
classification of symmetric spaces \cite{ford} and led to the infinite 
conservation laws of the model \cite{park}.

In this Letter, we consider the generalized Heisenberg ferromagnet model in 
(2+1)-dimensions. The motivations are twofold. Firstly, the model itself 
can be used in describing generalized planar ferromagnetisms where the 
generality coming from the large degrees of freedom of symmetric spaces can 
be used to handle various physical situations. 
Secondly, by gauging and coupling the model with the Chern-Simons gauge fields, 
we are led to the Chern-Simons self-dual solitons \cite{jackiw} which attracted  
an upsurge of recent interests in regard of the application to the quantum Hall
effect and the high-Temperature superconductivity \cite{pran,ston,wilz}.
Here, we focus on the second motivation and show that, using the properties of Hermitian 
symmetric spaces, the Hamiltonian of the model is bounded below by a topological 
charge. The resulting Chern-Simons solitons satisfy a vortex-type equation when 
the model is gauged with the maximal torus subgroup of $H$ and added by a gauge invariant 
term which induces vacuum symmetry breaking. To our knowledge, this vortex-type 
equation is new and is likely to possess similar properties to those of the 
vortex equation of Abelian Higgs model \cite{hong} or gauged non-linear
Schr\"odinger model \cite{jack}. As an example, we present an explicit 
expression for the vortex-type equation in the case of $CP(N)$.

We first recall that the action principle for the (1+1)-dimensional 
generalized Heisenberg ferromagnet model defined on the Hermitian symmetric 
space $G/H$ can be given by \cite{park}
\begin{equation}
S= \int dt dx [ \mbox{ Tr }(2Kg^{-1}\pp_t g+
\pp_{x}(gK g^{-1})\pp_{x} (gK g^{-1}))] 
\label{preact}
\end{equation}
where $g$ is a map $g:R^{1+1} \rightarrow G$ for the Lie group $G$ and $K$ 
is an element in the Cartan subalgebra of the Lie algebra ${\bf g} (= 
{\bf h} \oplus {\bf m} ) $ whose centralizer in ${\bf g}$ is ${\bf h}$, 
i.e. ${\bf h} = \{ V \in {\bf g} ~ : ~ [V ~ , ~ K] = 0 \}$.    
Up to a scaling, $J \equiv \mbox{ad}K = [K, ~ *]$ defines a linear map 
$J: {\bf m} \rightarrow {\bf m}$ which satisfies the complex structure
condition $J^{2} = -1 $ or, 
\be
[K, ~ [K,~ M]] = - M, 
\label{hsym}
\ee
for $ M \in {\bf m}$.
The action $S$ in Eq. (\ref{preact}) possesses local symmetry under $g \rightarrow 
gh$ for $h: R^{1+1} \rightarrow H$ so that the physical variable is  
effectively given by a generalized spin variable $Q \equiv gKg^{-1}$ valued in 
the coadjoint orbit of $G/H$. 

Now, we extend the model to the (2+1)-dimensional case and introduce gauge 
fields $A_{\m}$ which gauges the left multiplication of group $G ~ ; ~
g \rightarrow g^{'}g$. 
Consider the (2+1)-dimensional gauged Heisenberg ferromagnet model defined by the 
following action;
\begin{equation}
S= \int dt d^2x\left\{[\mbox{ Tr }(2Kg^{-1}D_t g+
D_i(gK g^{-1})D_i (gK g^{-1}))]-V(gKg^{-1})+
{\cal L_{CS}}\right\}.
\label{action}
\end{equation}
We use the convention in which the generators $T^A$'s
satisfy the commutation relation, $[T^A,T^B]=f^{ABC}~T^C$,
and the normalization, Tr$(T^AT^B)=(-1/2)\delta_{AB}$.
The covariant derivative is defined on fundamental and adjoint 
representations by
\begin{equation}
D_\mu g =\partial_\mu g +A_\mu g,~~~A_\mu=A_\mu^A T^A , ~~~ 
D_{\m }(gKg^{-1} ) = D_{\m }Q = [ \pp_{\m } + A_{\m } ~ , ~ Q] .
\end{equation}
The potential $V(gKg^{-1})$ is given in terms of
the generalized spin  $Q^A=-2\mbox{Tr}(QT^A)$:
\begin{equation}
V=\frac{1}{2}I^{AB}Q^AQ^B
\label{potential}
\end{equation}
where $I^{AB}$ is a constant symmetric matrix measuring the anisotropy of 
the system \cite{kose}. We assume that the dynamics of gauge fields is 
governed by the Chern-Simons action ${\cal L}_{CS}$:
\begin{equation}
{\cal L}_{CS}=-\kappa\epsilon^{\mu\nu\rho}\mbox{Tr}
(\partial_\mu A_\nu A_\rho+\frac{2}{3}A_\mu A_\nu A_\rho).
\end{equation}
Then, the equations of motion in terms of the generalized spin $Q$ are the 
gauged planar Landau-Lifshitz equation;
\begin{equation}
D_{t} Q+D_i [Q,~D_i Q]+[\bar Q,Q]=0
\label{hequation}
\end{equation}
and the gauge field equation;
\begin{equation}
\frac{\kappa}{2}\epsilon^{\mu\nu\rho}F_{\mu\nu}^A=J^{\mu A}
\end{equation}
where $\bar Q=I^{AB}Q^BT^A$.  The current density is given by
\begin{equation}
J^{ A}=(Q^A, -2\mbox{Tr}(T^A[Q,D^i Q]).
\end{equation}
In particular, the zeroth component gives the Gauss's law constraint:
\begin{equation}
G^A=  \frac{\kappa}{2}\epsilon^{ij}F_{ij}^A-Q^A =0
\end{equation}
We may rewrite the action Eq.(\ref{action}) in terms of Hamiltonian
\begin{equation}
S= \int dt d^2x\left\{\mbox{ Tr }(2Kg^{-1}\dot g)-\frac{\kappa}{2}
\epsilon^{ij}A^A_i \dot A^A_j-{\cal H}+A_0^AG^A\right\}
\end{equation}
where the Hamiltonian is given by
\be
H=\int d^2x {\cal H}=
\int d^{2}x [ \frac{1}{2}(D_{i}Q^A)^2 + V(Q^A) ] .
\ee
Owing to the property of Hermitian symmetric spaces Eq. (\ref{hsym}), we have
a useful identity;
\be
[Q, ~ [Q,~ D_i  Q]] =
 g[K, ~ [K, ~ [g^{-1} D_i  g, ~ K]]]g^{-1} 
= -g[g^{-1} D_{i}  g, ~ K]g^{-1} = - D_i  Q ,
\label{hermit}
\ee
which brings the Hamiltonian $H$ into the Bogomol'nyi type;
\ben
H &=& \int d^{2}x [\frac{1}{4}( D_{i}Q^A\pm\epsilon_{ij}
[Q,  D_{j}Q ]^A)^2 + V(Q^A) \pm {1 \over 2} 
\epsilon_{ij}Q^A [D_i Q ~ , ~ D_j Q]^A
 ]  \nonumber \\
&=& 
\int d^{2}x [\frac{1}{4}( D_{i}Q^A\pm\epsilon_{ij}
[Q,  D_{j}Q ]^A)^2 + V(Q^A) ]
\pm \frac{1}{2}\epsilon_{ij}F_{ij}^AQ^A ]
\pm 4\pi T .
\label{bogh}
\end{eqnarray}
The topological charge $T$ is defined by
\begin{equation}
T =\frac{1}{8\pi}
\int d^2x [\epsilon_{ij}Q^A 
[\partial_iQ,\partial_jQ]^A-2\epsilon_{ij}\partial_i (Q^AA_j^A)].
\label{tcharge}
\end{equation}
Thus, the energy is bounded below by the topological charge $T$ when the 
potential $V$ is chosen such that
\be
V \pm \frac{1}{2}\epsilon_{ij}F_{ij}^AQ^A = 0 .
\ee
Or, upon imposing the Gauss's law constraint, it is equivalent to 
choosing the constant matrix 
\begin{equation}
I^{AB}=\mp\frac{2}{\kappa}\delta_{AB} .
\end{equation}
The minimum energy arises when the spin variable satisfies the 
first order self-duality equation,
\begin{equation}
D_iQ=\mp \epsilon_{ij}[Q, D_jQ].
\label{selfdual}
\end{equation}
By taking a commutator with $Q$, we note that Eq. (\ref{selfdual}) 
is consistent with Eq. (\ref{hermit}).
For a given sign of the potential, the energy minimizing 
solutions are either self-dual or anti self-dual, but not both. 
As in the case of the non-relativistic non-linear Schr\"odinger 
model \cite{jack}, this is in  contrast with the Chern-Simons solitons
of the relativistic Abelian Higgs model \cite{hong} or the
gauged nonlinear sigma model \cite{kimm} where a fixed potential 
admits both self-dual and anti self-dual solitons as energy minimizing 
static configurations. The specific choice of the potential Eq. (\ref{potential})
describes isotropic case where the potential trivially reduces to a 
constant. However, if we choose the gauge group to be a proper subgroup of 
$G$, then the potential becomes nontrivial and describes anisotropic 
cases.
Another observation to be made is that in the absence of gauge fields 
the self-duality equation (\ref{selfdual}) is precisely 
that of two dimensional instantons in the principle chiral model which has been classified 
according to each symmetric spaces \cite{pere}. The role of the gauge field 
is that, through the Chern-Simons dynamics, it changes instantons into vortices. 

In order to see how vortices arise in our model, we take the gauge group  
to be the maximal torus subgroup of $H$ and introduce gauge invariant 
terms to the action which induce vacuum symmetry breaking. Explicitly, we
take $H^{a} (a = 1, \cdots , \mbox{rank} (H))$ to be generators of the maximal 
torus group and add to the action Eq. (\ref{action}) a linear term
\begin{equation}
\D S=\int dtd^{2}x A_o^av^a
\end{equation}
where each $v^a$ is a constant and the sum is taken over $a=1, \cdots, 
\mbox{rank} (H)$. Up to a total derivative term, $\D S$ is invariant under the gauge 
transform. 
Then, the gauge fields $A_{\m } = A_{\m}^{a}H^{a}$ and the Chern-Simons 
action reduces to a sum of Abelian Chern-Simons terms,
\begin{equation}
{\cal L}_{CS}=\frac{\kappa}{2}\epsilon^{\mu\nu\rho}
\partial_\mu A^a_\nu A^a_\rho ~ .
\end{equation}
The Gauss's law is replaced by
\begin{equation}
\frac{\kappa}{2}\epsilon_{ij}F^a_{ij} =Q^a-v^a.
\label{agauss}
\end{equation}
Also, we have the topological charge replacing Eq. (\ref{tcharge}) 
\begin{equation}
T=\frac{1}{8\pi}\int d^2x[\epsilon_{ij}Q^A [\partial_iQ, \partial_jQ]^A
+2\epsilon_{ij}\partial_i((v^a-Q^a)A^a_j)] .
\end{equation}
We assume the potential $V$ to be of the form
\begin{equation}
V(gKg^{-1})=\frac{1}{2}\sum_a I^a(Q^a-v^a)^2 ,
\label{pot2}
\end{equation}
then the Bogomol'nyi bound is established with the choice
\begin{equation}
I^1=\cdots =I^{N-1}=\mp\frac{2}{\kappa}.
\end{equation}
Note that the potential Eq. (\ref{pot2}) is nontrivial unlike the 
previous case and the nonvanishing constants $v^{a}$ breaks 
the symmetry of the vacuum spontaneously.

In the following, in order to be more explicit,  we restrict to the 
$CP(N-1) = SU(N)/(SU(N-1) \times U(1))$ case and present a detailed analysis.
In this case, the element $K$ in the Cartan subalgebra is given by 
$K=(i/N) \mbox{diag}(N-1,-1,\cdots, -1)$. Now introduce a parameterization of the 
group element $g$ of $SU(N)$ by an $N$-tuple, $ g = (Z_1, Z_2,\cdots, Z_N);
\ Z_p\in {\bf C}^N$ $(p,q=1,\cdots,N)$, such that
\begin{equation}
\bar Z_pZ_q=\delta_{pq}, \quad \mbox{det}
(Z_1, Z_2,\cdots, Z_N) =1.
\label{cond}
\end{equation}
Then the generalized spin $Q$ is given by
\begin{equation}
Q=iZ_1\bar Z_1-iI .
\label{isosp}
\end{equation}
All other $Z_p$'s with $p=2,\cdots,N$ disappear in the expression 
of $Q$ due to the particular form of $K$.
In terms of the Fubini-Study coordinate 
$\psi_\alpha (\alpha=1,2,\cdots, N-1$) \cite{alek}:
\begin{equation}
z_1=\frac{1}{\sqrt{1+\vert\psi\vert^2}}~ , ~ 
z_{\alpha +1}=\frac{\psi_\alpha}{\sqrt{1+\vert\psi\vert^2}},\quad
 ~;~ Z_1^T=(z_1,z_2,\cdots,z_N) ,
\label{slp}
\end{equation}
we have an equivalent expression of $Q$ in component, 
\begin{equation}
Q^A(\psi,\bar\psi)=-2i\sum_{p,q=1}^{N}\bar z_p(T^A)_{pq}z_q .
\label{spinfunction}
\end{equation}
We choose the standard expression for $T^A$'s:
$T^A=i\lambda^A/2$ where $\lambda^A$ is the  $SU(N)$ Gell-Mann
matrices \cite{geor}. The Cartan subalgebra generators $H^{a}$ 
generating the maximal torus group of $SU(N-1) \times U(1)$ 
are given by $N-1$ diagonal matrices 
\begin{equation}
H^a_{pq}=i(\sum_{k=1}^{a}\delta_{ik}\delta_{jk}
-a\delta_{i,a+1}\delta_{j,a+1})/\sqrt{2a(a+1)}~;~ a = 1, \cdots, N-1 .
\end{equation}
Using the complex notation;
 $z = x+iy, \bar z = x-iy$,
$A_z  = \frac{1}{2}(A_1 - iA_2), A_{\bar z} =
\frac{1}{2} (A_1 + iA_2)$,  and
$D_z = \frac{1}{2}(D_1-iD_2),
D_{\bar z} = \frac{1}{2}(D_1+iD_2)$, we obtain
an alternative expression of the self-duality equation, 
\begin{equation}
D_zQ=\mp i[Q, D_zQ].
\label{selfdualeq}
\end{equation}
With the parameterization of $Q$ as in Eq. (\ref{spinfunction}), the self-duality 
equation (\ref{selfdualeq}) for the plus sign case becomes a set of $N-1$ equations,
\begin{eqnarray}
(\partial_z &+&iA_z^1)\bar\psi_1=0 \nonumber\\
(\partial_z &+&\frac{i}{2}(A_z^1+{1 \over \sqrt{3}}A_z^2))\bar\psi_2=0\nonumber\\
(\partial_z &+&\frac{i}{2}(A_z^1+\frac{1}{\sqrt{3}}A_z^2
+\frac{4}{\sqrt{6}}A_z^3))\bar\psi_3=0\\
   & &\cdots\nonumber\\
(\partial_z &+&\frac{i}{2}(A_z^1+\frac{i}{\sqrt{3}}A_z^2
+\cdots +\frac{1}{\sqrt{(N-1)(N-2)/2}}A^{N-2}_z+
\frac{N}{\sqrt{N(N-1)/2}}A^{N-1}_z))
\bar\psi_{N-1}=0.\nonumber
\label{hiearchy}
\end{eqnarray}
Or, in terms of a shorthand notation
\begin{equation}
D_-^{\alpha } \equiv \partial_z+\frac{i}{2}(A_z^1+\frac{1}{\sqrt{3}}A_z^2
+\cdots +\frac{1}{\sqrt{(\alpha-1)(\alpha-2)/2}}A^{\alpha-2}_z+
\frac{\alpha}{\sqrt{\alpha(\alpha-1)/2}}A^{\alpha-1}_z)
\end{equation}
we have
\begin{equation}
D_-^\alpha\bar\psi_\alpha=0 ~ ; ~ \a = 1, \cdots , N-1.
\label{equa1}
\end{equation}
Similarly, for the minus sign case, we have
\begin{equation}
D_+^\alpha \equiv \partial_z-\frac{i}{2}(A_z^1+\frac{1}{\sqrt{3}}A_z^2
+\cdots +\frac{1}{\sqrt{(\alpha-1)(\alpha-2)/2}}A^{\alpha-2}_z+
\frac{\alpha}{\sqrt{\alpha(\alpha-1)/2}}A^{\alpha-1}_z)
\end{equation}
and
\begin{equation}
D_+^\alpha\psi_\alpha=0 ~ ; ~ \a = 1, \cdots , N-1.
\label{equa2}
\end{equation}
The Gauss's law constraint Eq. (\ref{agauss}) is given by
\begin{equation}
\partial_z A^a_{\bar z}-\partial_{\bar z}A^a_z=
Q^a(\psi,\bar\psi)-v^a.
\end{equation}

In the $CP(1)$ case, we have only one complex $\psi $ which we parameterize by
\be
\bar\psi=\rho\exp(i\phi)
\ee
where $\rho $ is real and the phase $\phi $ is a real multi-valued function.
Then, Eq. (\ref{equa1}) can be solved for the gauge field $A$  and
Eq. (\ref{agauss}) reduces to a vortex-type equation;
\begin{eqnarray}
A_i&=& \epsilon_{ij}\partial_j\log\rho-\partial_i\phi
\nonumber\\
\nabla^2 \log\rho+\epsilon_{ij}\partial_i\partial_j\phi
&=& \frac{1}{\kappa}(v-\frac{1-\rho^2}{1+\rho^2}).
\end{eqnarray}
The derivative term $ \epsilon_{ij}\partial_i\partial_j\phi $ is identically 
zero except at the zeros of $\bar\psi $ where the multi-valuedness of $\phi $ 
results in the Dirac delta function(see, for example, \cite{taubes}).
In the $CP(2)$ case, we define 
$\bar\psi_1=\rho_1\exp(\phi_1),~\bar\psi_2=\rho_2\exp(\phi_2) $ and solve 
Eq. (\ref{equa1}) for $A^{1}$ and $A^{2}$ so that the resulting vortex-type 
equation becomes
\begin{eqnarray}
A^1_i&=&\epsilon_{ij}\partial_j\log\rho_1-\partial_i\phi_1
\nonumber\\
A^2_i&=&\frac{2}{\sqrt{3}}\epsilon_{ij}\partial_j\log\rho_2
-\frac{1}{2}\epsilon_{ij}\partial_j\log\rho_1-
\frac{2}{\sqrt{3}}\partial_i\phi_2+\frac{1}{2}\partial_i\phi_1
\nonumber\\
\nabla^2 \log\rho_1+\epsilon_{ij}\partial_i\partial_j\phi_1
&=& \frac{1}{\kappa}(v^1-\frac{1-\rho_1^2}{1+\rho_1^2+\rho_2^2})\\
\frac{2}{\sqrt{3}}\nabla^2\log\rho_2+\frac{2}{\sqrt{3}}\epsilon_{ij}
\partial_i\partial_j\phi_2
&=&\frac{1}{\kappa}(\frac{v^1}{2}+v^2-\frac{1}{2}
\frac{1-\rho_1^2}{1+\rho_1^2+\rho_2^2}-
\frac{1}{\sqrt{3}}\frac{1+\rho_1^2-2\rho_2^2}
{1+\rho_1^2+\rho_2^2}).\nonumber
\end{eqnarray}                              
In general for $CP(N-1)$, the number of $\psi$'s is half the degrees of 
freedom of $CP(N-1)$, that is, $(N^{2} - 1 - (N-1)^2 )/2= N-1$, so that 
Eq. (\ref{equa1}) can be solved for $A^{a} ~; ~ a= 1, \cdots , N-1$ and the 
Gauss's law reduces to the $N-1$ vortex-type partial differential 
equations. To our knowledge, these are new vortex-type equations.
A numerical analysis suggests that these vortex-type equations indeed 
possess vortex solutions and show a rich structure depending on the 
value of $v^{a}$'s. Details on the rotationally symmetric
solutions and their properties will appear elsewhere \cite{oh}. 
Here, we only contend that such vortices saturate the bound 
\begin{equation}
E=4\pi  \vert T\vert .
\end{equation}
The energy $E$ and the topological charge $T$ can be computed from 
the Hamiltonian 
\begin{equation}
E=4\int d^2x g_{\alpha\beta}(D_+^\alpha\psi_\alpha
\bar D_+^\beta \bar\psi_\beta+\bar D_-^\alpha\psi_\alpha
D_-^{\b }\bar\psi_\beta)
\end{equation}
where $g_{\alpha\beta}$ is the Fubini-Study metric on $CP(N-1)$,
\begin{equation}
g_{\alpha\beta}=\frac{(1+\vert\psi\vert^2)\delta_{\alpha\beta}
-\bar\psi_\alpha\psi_\beta}{(1+\vert\psi\vert^2)^2} , 
\end{equation}
and the expression for the topological charge in this case is
\begin{equation}
T ={1\over \pi }
\int d^2x \left\{ g_{\alpha\beta}(\partial\psi_\alpha\bar\partial\bar\psi_\beta
-\bar\partial\psi_\alpha\partial\bar\psi_\beta)
 +\frac{1}{4}\epsilon_{ij}\partial_i\left((v^a-Q^a(\psi,\bar\psi))A^a_j\right)
\right\} .
\end{equation}
In general, with an appropriate normalization, this topological charge $T$ is taken 
to be integer valued \cite{pere}.
In the absence of gauge fields, solutions
of Eq. (\ref{equa1}) and Eq. (\ref{equa2}) are simply given by
holomorphic or antiholomorphic functions,
$\psi_\alpha\equiv \psi_\alpha(z)$ or
$\psi_\beta\equiv \psi_\beta(\bar z)$, and the solutions of the
self-dual equation (\ref{selfdualeq}) is obtained by substituting
these functions into the Eq. (\ref{spinfunction}).
This generalizes the Belavin-Polyakov solution for the $CP(1)$ case
\cite{pere}. 

In conclusion, we have obtained a new vortex-type equations by 
considering the (2+1)-dimensional gauged generalized Heisenberg 
ferromagnet model coupled with the Chern-Simons gauge fields.
It was shown that the Hermitian symmetric space plays an essential 
role in deriving the Bogomol'nyi bound in addition to the specific  
choice potential. Our approach allows a systematic generalization of 
the vortex equation according to each symmetric spaces. 
It would be an interesting problem to extend our analysis on the  
vortices to other Hermitian symmetric space $G/H$, 
and study the properties of vortex-type nonlinear equations, e.g. existence  
of multi-vortices, in the general  
context of group theory.

\vglue .4in
{\bf ACKNOWLEDGEMENT}
\vglue .2in
PO would like to thank Yoonbai Kim for useful discussions.
This work is supported  by the program of Basic Science Research,
Ministry of Education  BSRI-96-1419/BSRI-96-2442 , and
by Korea Science and Engineering Foundation through the Center for
Theoretical Physics, SNU and through the project number 96-0702-04-01-3.

\end{document}